\documentclass[pra,aps,10pt,superscriptaddress,twocolumn,floatfix]{revtex4-1}
\usepackage{graphicx,color}
\usepackage[nice]{nicefrac}
\usepackage{amsmath,amssymb,bm}
\usepackage{amsmath}
\usepackage{braket}
\usepackage{float}
\usepackage{amssymb,amsmath,amsthm,enumitem}

\usepackage[plainpages=false,pdfpagelabels,colorlinks=true,linkcolor=red,urlcolor=blue,citecolor=blue,pdftitle={},pdfauthor={},pdfdisplaydoctitle=true,pdfduplex=DuplexFlipLongEdge]{hyperref}

\definecolor{darkred}{rgb}{0.90,0.2,0.2}
\definecolor{darkgreen}{rgb}{0,0.60,.2}
\definecolor{darkblue}{rgb}{0.1,0.3,1}
\definecolor{grey}{cmyk}{0,0,0,0.25}
\definecolor{orange}{cmyk}{0,0.6,0.8,0}

\begin{document}

\title{Long-lived non-thermal states in pumped one-dimensional systems of hard-core bosons}

\author{Patrycja {\L}yd{\. z}ba}
\affiliation{Department of Theoretical Physics, J. Stefan Institute, SI-1000 Ljubljana, Slovenia}
\affiliation{Department of Theoretical Physics, Wroclaw University of Science and Technology, 50-370 Wroc{\l}aw, Poland}
\author{Janez Bon{\v c}a}
\affiliation{Department of Theoretical Physics, J. Stefan Institute, SI-1000 Ljubljana, Slovenia}
\affiliation{Department of Physics, Faculty of Mathematics and Physics, University of Ljubljana, SI-1000 Ljubljana, Slovenia}

\begin{abstract}
We study a unitary time evolution of a symmetry-broken state in a form of a charge density wave in a finite system of interacting hard-core bosons, which can be mapped onto the XXZ Heisenberg chain. Moreover, we introduce a spatially-homogenous and time-dependent vector potential that mimics a short laser
pulse. We establish the range of amplitudes of the vector potential for which the onset of charge density wave order can be controlled. We propose a protocol that reveals non-thermal long-lived states, which are characterized by a non-zero charge density wave order translated by one lattice site with respect to its initial formation. The life times of these states are large in comparison to all typical times given by the parameters of the system. They increase with the number of lattice sites, but are significantly suppressed by the integrablility breaking perturbations. In view of these findings, we speculate that the long-lived non-thermal states exist in the thermodynamic limit.
\end{abstract}

\maketitle

\section{Introduction}

The non-equilibrium dynamics of isolated quantum systems has become a subject of interest in recent years. For example, the transfer of energy between different degrees of freedom has been analytically and numerically studied in Hubbard, Holstein and other models \cite{Mendoza_2017,Balzer_2015,Filippis_2012,Golez_2012}. Understanding the pathway of charge, spin and lattice excitations, together with determining their lifetimes, is also targeted by pump-probe experiments \cite{Alexandrou_1995,Pug_2007,Amo_2010,Nie_2015}. Driving the system out of equilibrium by a laser has an additional advantage, it offers a possibility to design states of matter, like coherent excitations and $\eta$-paired states in Mott antiferromagnetic insulators \cite{Wang_2017,Kaneko_2019}, which only appear under the influence of a light pulse. In the simplest protocol that realizes the non-equilibrium dynamics of an isolated quantum system, which is known as the quantum quench \cite{Rigol_2009b}, the system is prepared in a pure state $\ket{\Psi\left(0\right)}$ and allowed to unitarily evolve in time $\ket{\Psi\left(\tau\right)} = e^{-i\hat{H}\tau/\hbar} \ket{\Psi\left(0\right)}$. Naturally, $\ket{\Psi\left(0\right)}$ is not an eigenstate of $\hat{H}$. Its experimental implementation has been enabled by the development of ultra-cold quantum gases setups, which are almost perfectly isolated from the environment and their internal parameters can be controlled with a high precision \cite{Mazurenko_2017,Trotzky_2012,Trotzky_2008,Hofferberth_2007,Kinoshita_2006,Kinoshita_2004}. Among others, the inter-atomic interactions can be tuned with the help of an external magnetic field and the Feshbach resonance \cite{Chin_2010}. When the quantum quench protocol is realized in these experiments, the system is initially prepared in its ground state and then some internal parameters are abruptly changed \cite{Trotzky_2008,Trotzky_2012}. It is also worth to mention that other protocols, like inhomogeneous quenches in which geometric constraints imposed on a system are abruptly removed \cite{Luca_2017,Cook_2020}, open Markovian setups in which edges of a spin chain are kept in contact with spin baths \cite{Prosen_2011} or periodic quenches with certain parameters of a Hamiltonian in the form of a square wave \cite{Claeys_2017,Mishra_2019} can also be considered.

The main focus of studies of non-equilibrium dynamics has been on the description of relaxation towards steady-like states \cite{Reimann_2016}. It has been recognized that the majority of systems evolve in time towards states that appear to be in thermal equilibrium. Specifically, their long-time expectation values of observables agree with predictions of a grand canonical ensemble with a temperature and chemical potential fixed by an average energy and number of particles in the initial state, respectively. The relaxation of these so-called quantum-chaotic systems is believed to be described by the Eigenstate Thermalization Hypothesis \cite{Jansen_2019,Alessio_2016,Srednicki_1994,Deutsch_1991}. On the contrary, it has been found that the so-called integrable systems, which are characterized by an extensive amount of local conserved quantities \cite{Grabowski_1995,Mierzejewski_2015,Ilievski_2016}, can behave in a vastly different manner when taken far from equilibrium \cite{Rigol_2007,Caux_2013}. The long-time expectation values of observables can support non-vanishing oscillations with averages described by the Generalized Gibbs Ensemble \cite{Rigol_2009,Cassidy_2011,Ilievski_2015,Essler_2016,Vidmar_2016}, in which not only average energy and particle number but also other constants of motion are fixed. It should be mentioned that the residues of non-generic dynamics can be found in a system displaced from its integrability point. For example, the existence of approximate constants of motion can trap the system in long-lived non-thermal states, the phenomenon which is known as the prethermalization \cite{Kollar_2011,Moeckel_2008}. Additionally, the off-diagonal elements of the single-particle density matrix can sustain their oscillatory relaxation as presented for the infinite-dimensional Fermi-Hubbard model on Bethe lattice \cite{Balzer_2015}.

It is not always easy to design a protocol in which some arbitrary parameter of a steady-like state clearly deviates from its thermal prediction. Let us consider a widely-studied scenario, i.e., the unitary time evolution of the N{\'e}el state in an exactly solvable interacting integrable model \cite{Woulters_2014,Pozsgay_2014,Fagotti_2013,Fagotti_2014}. As established for the XXZ Heisenberg chain in \cite{Barmettler_2009,Barmettler_2010,Collura_2020}, the staggered magnetization exponentially decays to zero for any finite anisotropy parameter $\Delta$, while the relaxation time diverges as $\tau_r\propto\log\Delta$ when $\Delta\rightarrow 0$, and $\tau_r\propto\Delta^2$ when $\Delta\rightarrow\infty$. There are additional oscillations of the order parameter in the short-time relaxation in the gapless phase $\Delta<1$, which are absent in the gapped phase $\Delta>1$. On the other hand, the existence of additional quasi-local conserved charges in the gapless phase, which become non-local when $\Delta\rightarrow 1$, is manifested by the appearance of a persistent current after a flux quench, i.e., an excitation by electric field with a Dirac delta profile \cite{Nakagawa_2016,Luca_2017}. The junction of two states with non-vanishing currents supports the formation of an expanding magnetic domain \cite{Luca_2017}. Furthermore, the integrability enables a fully reversible dynamics in the presence of a slowly varying flux, despite the absence of a gap for $\Delta>1$ \cite{Bastianello_2019}. It is worth to mention that the quasi-momentum distribution and the structure factor for density-density correlations are experimentally accessible and usually efficient in revealing the breakdown of thermalization \cite{Rigol_2009}.

In the paper we expand this widely-studied scenario by the inclusion of a short laser pulse with a frequency $\omega$ and a Gaussian profile in the gapped phase. We present our results in the hard-core boson (HCB) picture which can be mapped onto the spin picture. This mapping connects the charge density wave state with the N{\'e}el state. We demonstrate that the order parameter $\eta$ in a finite system does not vanish and can be dynamically controlled for small amplitudes of electric field $E_0<\frac{\hbar\omega}{qa}$, while it is significantly diminished during the pulse duration for moderate and large amplitudes of electric field, $E_0\gtrsim \frac{\hbar\omega}{qa}$. Parameters $q$ and $a$ correspond to the particle charge and lattice spacing, respectively. Furthermore, we propose a setup for which long-lived non-thermal states are established. They are characterized by a non-zero charge density wave order translated by one lattice site with respect to the original position, i.e., they are identified by a persistent negative value of $\eta$. Furthermore, the relaxation times of these long-lived non-thermal states increase with the number of lattice sites, and are strongly suppressed after the inclusion of integrability breaking terms in the Hamiltonian. Therefore, we expect them to exist in the thermodynamic limit. Whether the charge density wave order eventually decays to zero or not, remains an open question. Note that the long-lived non-thermal states are not related to the prethermalized states, since the considered system is integrable before and after the excitation by the laser pulse. They are rather somewhat related to the coherence states (i.e., the quasi-condensed states) that have been found in a one-dimensional system of interacting HCBs when a state -- in which all sites in the centre of a system are filled while all sites near the edges of a system are empty -- experiences a sudden quench of a confining potential and undergoes a sudden expansion along a one-dimensional chain \cite{Vidmar_2015_hcb}.

\section{Model}

We consider a finite one-dimensional system of HCBs with nearest-neighbour interactions subjected to a spatially-homogeneous time-dependent vector potential that mimics a laser pulse
\begin{equation}
\mathcal{A}\left(\tau\right) = Ae^{-\frac{\left(\tau-\tau_0\right)^2}{2\sigma^2}}
\cos\left(\omega\left(\tau-\tau_0\right)\right)
\end{equation}
and enters the Hamiltonian via the Peierls' subsitution,
\begin{equation}
\label{eq1}
\mathcal{H}\left(\tau\right) = -t\sum_{j=1}^{L}\left(e^{i\mathcal{A}\left(\tau\right)}
c^{\dagger}_{j}c_{j+1}+\text{H.c.}\right)+V\sum_{j=1}^{L}n_{j}n_{j+1}.
\end{equation}
Here, $c^{\dagger}_{j}$ ($c_{j}$) creates (annihilates) a HCB at site $j$, while $n_{j} = c^{\dagger}_{j}c_{j}$. The commutation relations are as follows $\left[c_i,c_j^\dagger\right]=\delta_{i,j}\left(1-2n_i\right)$. The parameter $t$ is the nearest-neighbour hopping amplitude and $V$ is the strength of nearest-neighbour repulsive interactions. Furthermore, $A$, $\sigma$ and $\omega$ correspond to the amplitude, duration and frequency of a laser pulse, respectively. For a convenience, we set $\hbar$, $t$, $q$ and $a$ to unity, so that energy is expressed in units of $t$, time $\tau$ in units of $\frac{\hbar}{t}$ and amplitude $A$ in units of $\frac{\hbar}{qa}$.

We perform a quantum mechanical time evolution of a finite one-dimensional system of HCBs with Hamiltonian from Eq.~\ref{eq1} using the well-known Lanczos algorithm \cite{Park_1986,Ojalvo_1970,Lanczos_1950}. We chose the initial state to be a chain of alternately filled and unfilled sites $\ket{\Psi\left(0\right)} = \prod_{j=1}^{L/2} c_{2j-1}^\dagger\ket{\emptyset}$, which can be prepared with ultracold atoms in optical superlattices \cite{Trotzky_2008,Trotzky_2012}. Furthermore, we primarly focus on the time evolution of the order-parameter
\begin{equation}
\eta\left(\tau\right) = \frac{2}{L}\sum_{j=1}^{L} \left(-1\right)^{j} \bra{\Psi\left(\tau\right)}  c^{\dagger}_{j}c_{j} \ket{\Psi\left(\tau\right)}.
\end{equation}
It should be emphasized that the Jordan-Wigner transformation maps the investigated model onto the XXZ Heisenberg chain with coupling $J=2t=2$ and anisotropy parameter $\Delta=V/2$, as well as transforms $\ket{\Psi\left(0\right)}$ into the antiferromagnetic N{\'e}el state, i.e., the ground state in the limit $\Delta\rightarrow\infty$. The connection with the XXZ Heisenberg chain becomes invalid when next-nearest-neighbour terms are included in the Hamiltonian from Eq.~\ref{eq1}, and the particle statistics becomes relevant \cite{Barmettler_2010}. The investigated model experiences a quantum phase transition from a gapless to a gapped ground state for a critical interaction strength $V=2$. Moreover, this quantum phase transition is reflected in the time evolution of $\ket{\Psi\left(0\right)}$, i.e., $\eta\left(\tau\right)$ is described by the zeroth-order Bessel function with a decaying amplitude for $V<2$, and by the exponential function for $V>2$ \cite{Barmettler_2009,Barmettler_2010}.

We restrict considerations to the strongly correlated regime $V\gg 2$. In the weakly correlated regime, the melting of the order parameter is almost unaffected by the laser pulse and $\eta\left(\tau\right)\rightarrow 0$ for $\tau\rightarrow\infty$. Unless otherwise specified, we fix the number of sites to $L=24$, the interaction strength to $V=7$, and we introduce laser pulses that are symmetric around $\tau_0=6.0$. Furthermore, we consider durations $\sigma\in\left[0.3,...,5.0\right]$, for which electric field vanishes when $\tau\rightarrow 0$ and its Fourier transform has a peak near $\omega$. It should be mentioned that we cannot take advantage of the translational invariance to calculate the order parameter, and we need to evolve the system for a long time $\tau\leq 50$ to determine relaxation times of long-lived non-thermal states (we show evolutions with $\tau<50$ in some figures for clarity). Furthermore, we choose the time step to be $\delta t = 10^{-2}$ before and during the laser pulse, while $\delta t = 10^{-1}$ after the laser pulse to speed up the calculations. This limits the number of lattice sites that we can consider to $L\leq 26$.

As pointed out in the introduction, the investigated model is integrable before and after the excitation by a laser pulse \cite{Grabowski_1995}. The integrability can be broken either by the inclusion of the next-nearest-neighbour hopping $t^{'} \sum_{j=1}^L \left(e^{i2A\left(\tau\right)} c_j^\dagger c_{j+2} + \text{H.c.} \right)$ or interactions $V^{'}\sum_{j=1}^{L} n_j n_{j+2}$ \cite{Santos_2010}. We take advantage of this integrability breaking later in the paper.

\section{Results: Vanishing of order parameter}
\begin{figure}[t]
\centering
\includegraphics[width=0.8\columnwidth]{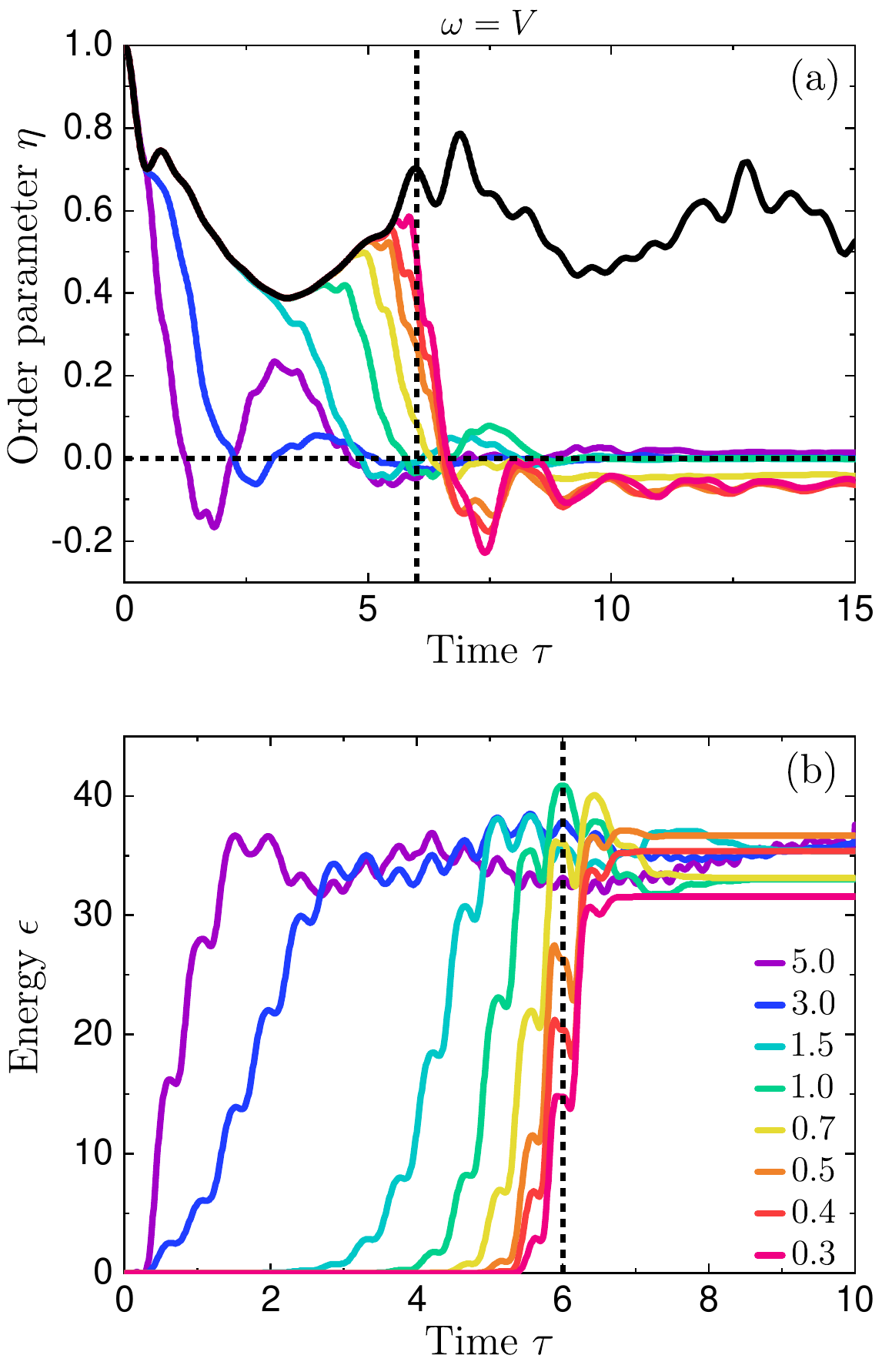}
\caption{Resonant-like pumping with $A=2.5$ and $\omega=V$. (a) The time evolution of the order parameter for selected pulses with durations $\sigma\in\left[0.3,...,5\right]$. The black reference curve corresponds to the time evolution without pumping. Note that it never decays to zero. Additionally, the horizontal dashed line marks $\eta=0$, while the vertical dashed line marks $\tau_0$. (b) The energy of a laser pulse absorbed by HCBs $\epsilon=\bra{\Psi\left(\tau\right)} \mathcal{H}\left(\tau\right) \ket{\Psi\left(\tau\right)}$. The legend presented in (b) is valid for (a) as well.}
\label{fig1}
\end{figure}
\begin{figure}[t]
\centering
\includegraphics[width=0.8\columnwidth]{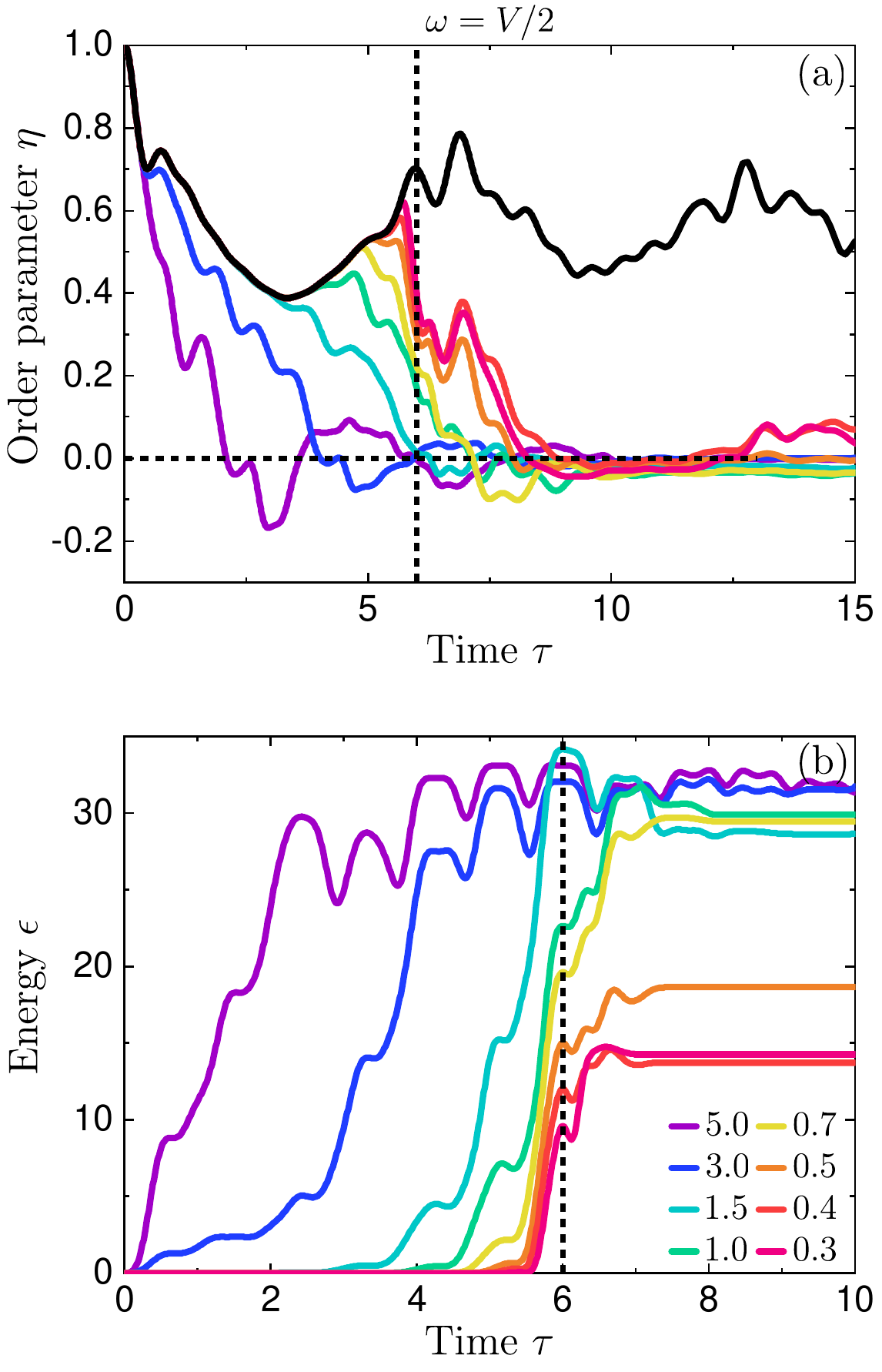}
\caption{Resonant-like pumping with $A=2.5$ and $\omega=V/2$. (a) The time evolution of the order parameter for selected pulses with durations $\sigma\in\left[0.3,...,5\right]$. The black reference curve corresponds to the time evolution without pumping. Note that it never decays to zero. Additionally, the horizontal dashed line marks $\eta=0$, while the vertical dashed line marks $\tau_0$. (b) The energy of a laser pulse absorbed by HCBs $\epsilon=\bra{\Psi\left(\tau\right)}  \mathcal{H}\left(\tau\right) \ket{\Psi\left(\tau\right)}$. The legend presented in (b) is valid for (a) as well.}
\label{fig2}
\end{figure}
\begin{figure}[t]
\centering
\includegraphics[width=0.8\columnwidth]{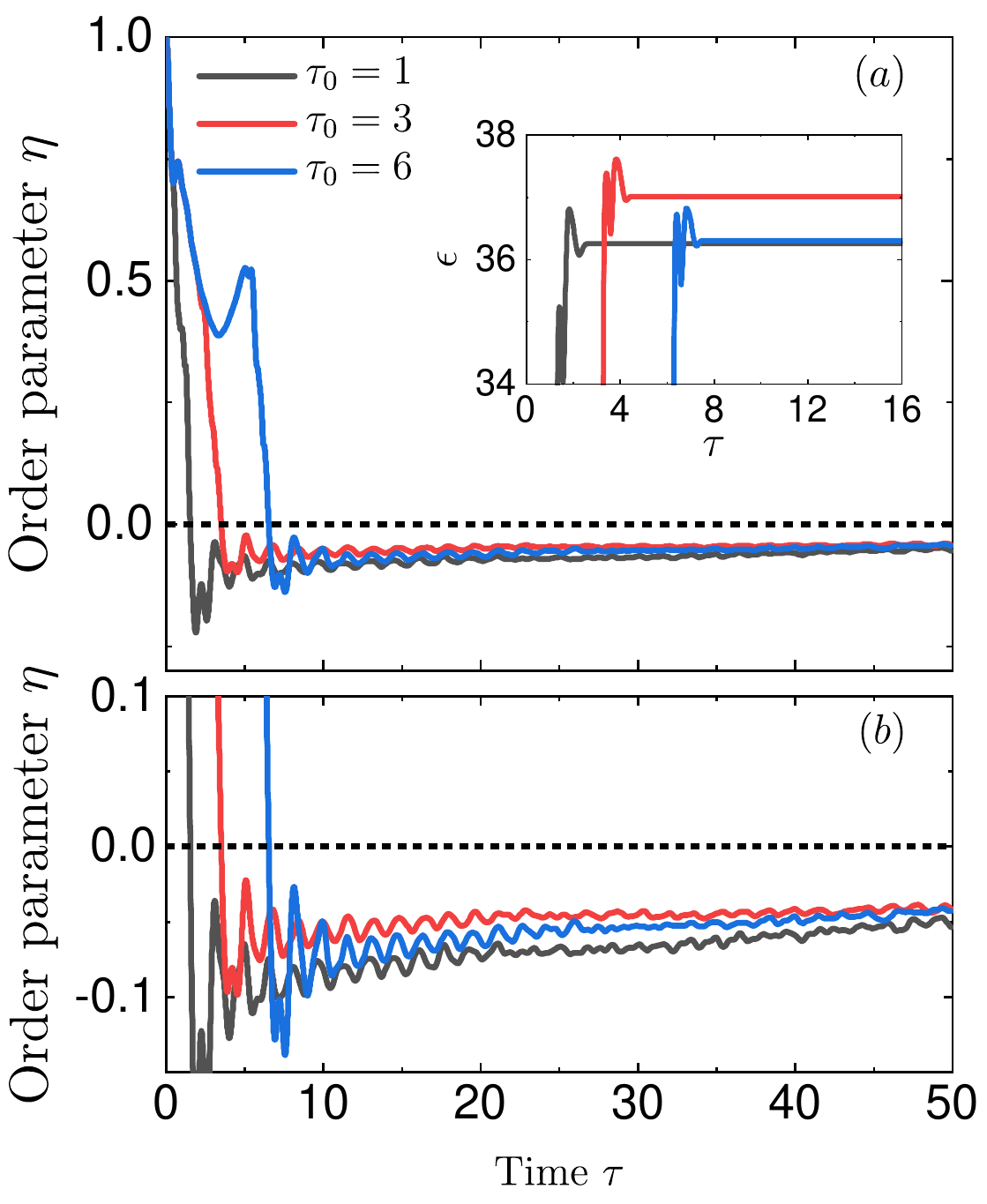}
\caption{Time evolution of $\eta$ (a) and $\epsilon$ (inset) for a laser pulse with $A=2.5$, $\omega = V$ and $\sigma=0.5$. Three different times $\tau_0=6, 3$ and $1$ have been presented. The dashed horizontal line marks $\eta = 0$. The order parameter of the steady-like state is negative independent of $\tau_0$. Simultaneously, the absorbed energy fluctuates with $\tau_0$. The same time evolutions as in (a) are presented in (b) but on a restricted scale $\eta\in[-0.15,0.1]$. This reveals that the long-time value of the order parameter and its progression towards zero also depend on $\tau_0$. }
\label{figR1}
\end{figure}

We have examined two examples of a resonant-like pumping with $\omega=V$ and $\omega=V/2$. Nevertheless, we have established that for a slight detuning from a resonance, $\omega\rightarrow\omega+\delta\omega$ with $\delta\omega\lesssim 1$, the presented results remain qualitatively unchanged. This is desirable from the experimental point of view but is not unexpected, since the Fourier transform of the electric field is non-zero for frequencies close to $\omega$. The order parameter is presented in Fig.~\ref{fig1}(a) for $\omega=V$, and Fig.~\ref{fig2}(a) for $\omega=V/2$. The energy absorbed by HCBs is depicted in Fig.~\ref{fig1}(b) for $\omega=V$, and Fig.~\ref{fig2}(b) for $\omega=V/2$. The time evolutions have been calculated for $A=2.5$, but they are qualitatively similar for all amplitudes in the range $A\gtrsim 1$.

The energy absorbed by HCBs
\begin{equation}
\epsilon=\bra{\Psi\left(\tau\right)} \mathcal{H}\left(\tau\right) \ket{\Psi\left(\tau\right)}
\end{equation}
is not proportional to the duration of a laser pulse even though the transmitted energy is linear in $\sigma$
\begin{equation}
\label{eqE}
E\approx A^2\omega^2\int_{-\infty}^{\infty} e^{-\frac{\left(\tau-\tau_0\right)^2}{\sigma^2}} \sin^2\left(\omega\left(\tau-\tau_0\right)\right) d\tau \propto A^2\omega^2\sigma.
\end{equation}
Nevertheless, the resonant-like excitation is not a weak perturbation, and the system response may not be described by the linear response theory. It actually seems that $\epsilon$ depends on $\sigma$ as well as on the state $\Psi\left(\tau\approx\tau_0-2\sigma\right)$ that begins to experience the vector potential of a pulse. The latter statement is confirmed in Fig.~\ref{figR1}, which presents how the time evolution of the order parameter and absorbed energy changes when the pulse with $A=2.5$, $\omega = V$ and $\sigma=0.5$ is introduced to the system at different times $\tau_0=6,3$ and $1$. It is clearly visible that the order parameter of the steady-like state remains negative to the longest times performed in our calculations independent of $\tau_0$ (a thorough analysis of the value of $\eta$ in the infinite-time limit is presented in the next section). Simultaneously, the absorbed energy fluctuates with $\tau_0$.

Consider the first scenario with $\omega = V$, as shown in Fig. \ref{fig1}(b). For all durations, the energy transmitted by a pulse, from Eq. (\ref{eqE}), is greater than the maximal energy HCBs can absorb, $V\sum_{j=1}^{L} \braket{n_{i}n_{i+1}}=V L/4=42$, which can be estimated from the atomic limit in which HCBs form nearest-neighbour pairs. In general, the response of a system appears to satisfy the following relations. When pulses are short $\sigma<1$, the steady-like state is described by a negative order parameter. Such a state represents lower potential energy than a state with zero order parameter. Furthermore, the absorbed energy in most cases increases with a pulse duration. When pulses are long $\sigma>1$, the steady-like state is described by zero order parameter. Again, the absorbed energy in most cases increases with a pulse duration up to $\sigma=3$. Note that for $\sigma>3$, HCBs experience the vector potential of a pulse from the beginning of the time evolution. Therefore, the saturation observed for $\sigma>3$ is due to the same state $\Psi\left(\tau\approx\tau_0-2\sigma\right)\approx\Psi\left(0\right)$. The saturated energy is only about $13\%$ lower than the maximal energy HCBs can absorb. 

In the second scenario with $\sigma=V/2$, the transmitted energy is lower than in the first scenario with  $\omega = V$, and the absorption requires the accompaniment of higher-order effects. The steady-like states with negative order parameters are not observed. Furthermore, the absorbed energy $\epsilon$ initially increases with the pulse duration $\sigma$, and then saturates near $\sigma=3$.

Note that the considered system is not in its energy eigenstate for $\tau=0$. It evolved in time before experiencing the vector potential of a laser pulse. From this point of view, it seems less surprising that the response, e.g., the absorbed energy, depends on the time when the laser pulse starts acting on the system, i.e., $\tau\approx\tau_0-2\sigma$. This observation can be important for pump-probe-like experiments.

\begin{figure}
\centering
\includegraphics[width=0.8\columnwidth]{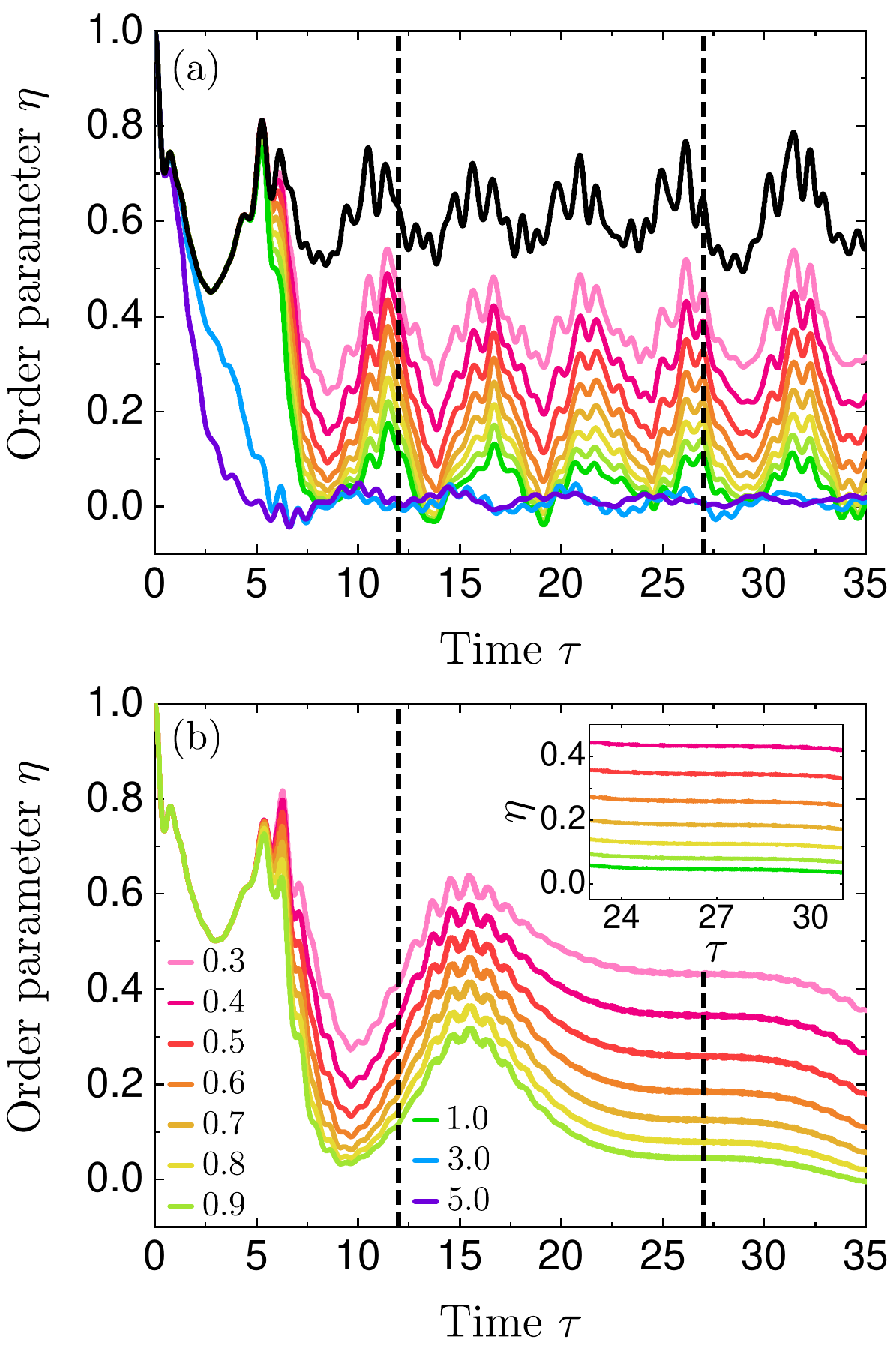}
\caption{(a) Resonant-like pumping with $A=0.5$ and $\omega=V$. A system with $L=20$ is considered. The time evolution of the order parameter for exemplary pulses with $\sigma\in\left[0.3,...,5\right]$ is presented. Note that the order parameter never decays to zero for $\sigma<1.0$. Instead, it oscillates around a time-independent average $\bar{\eta}> 0$, which decreases with a pulse duration. The black reference curve establishes $\eta\left(\tau\right)$ without pumping. (b) The time evolution of the order parameter for the protocol described in the main text. The one-dimensional system is initially excited with the same pulse as in (a), and then with a second pulse with $A=0.5$, $\omega=95.0$ and $\sigma=15.0$. The latter freezes the dynamics of HCBs. This protocol allows to create a plateau in the time evolution (see the inset in (b) for a close-up) with the value and size controlled by the duration of first and second pulse, respectively. The right dashed line marks $\tau_0=27$, while the left dashed line marks $\tau_0-\sigma=12$ of the second pulse. The legend presented in (b) is valid for (a) as well.}
\label{fig3}
\end{figure}

\begin{figure*}[t]
\centering
\includegraphics[width=0.9\textwidth]{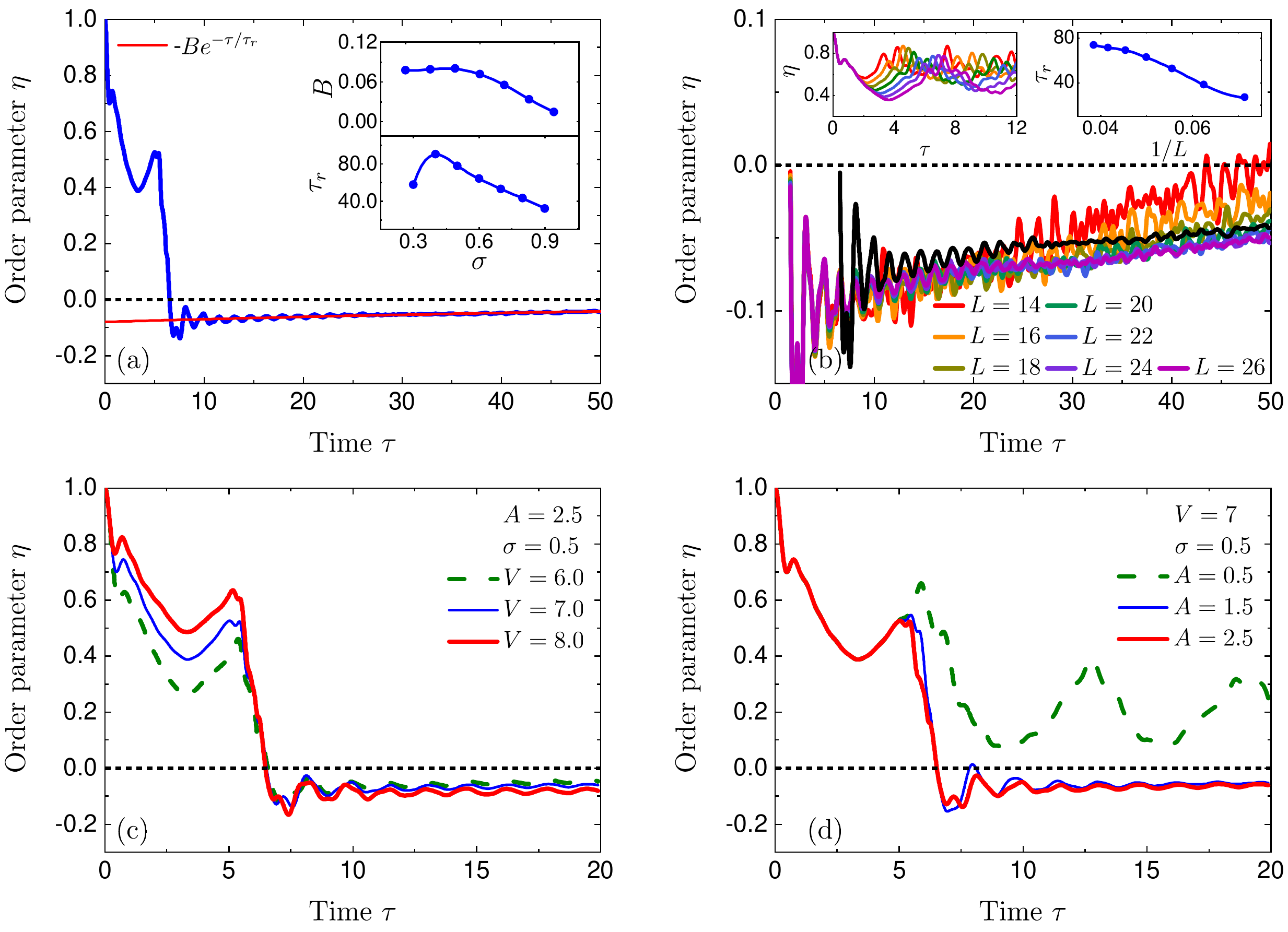}
\caption{(a) The time evolution of the order parameter after pumping with $A=2.5$, $\omega=V$ and $\sigma=0.5$. The non-thermal $\eta\left(\tau\right)<0$ is maintained for long times $\tau> 50$. Its gradual progression towards zero is satisfyingly described by $\eta\left(\tau\right)=-B e^{-\tau/\tau_r}$, and marked by the thin red line. The inset displays the dependence of $B$ and $\tau_r$ on $\sigma$. The continuous curves are cubic splines which are guides to the eye. (b) The comparison of long-lived non-thermal states for different system sizes $L\in\left\{14,16,...,26\right\}$. The same pumping protocol as in (a) is used except for a different value of $\tau_0=1$, so that the system is excited with a laser pulse from the beginning of the evolution, and we can neglect the fact that the revival of nearest-neighbour correlations occurs at different times for different system sizes. The black curve marks the long-lived non-thermal state for a system with $L=24$ and an excitation at $\tau_0=6$. The right inset shows how the relaxation time increases with the number of lattice sites $L\in\left\{14,16,...,26\right\}$. The continous curve is a cubic spline. The left inset shows the time evolution of the order parameter without pumping. Figures (c) and (d) investigate the stability of a long-lived non-thermal state form (a) against changes in the interaction strength $V$ and amplitude $A$, respectively.}
\label{fig4}
\end{figure*}

An important observation regarding the time evolution of the order parameter is that $\eta\left(\tau\right)$ does not decay to zero in a finite system unperturbed by a laser pulse (see the black reference curve in Fig.~\ref{fig1}(a) or Fig.~\ref{fig2}(a)). Instead, the order parameter decreases for short times $\tau\leq \tau_\text{rev}\approx 4.0$ and then oscillates around a time-independent average $\bar{\eta}\approx 0.6$. The time of revival of nearest-neighbour correlations  (i.e., the time of appearance of oscillations $\tau_\text{rev}$) increases with the number of lattice sites as visible in the inset of Fig.~\ref{fig4}(b), which is consistent with the exponentially-slow decay of the order parameter to zero in the thermodynamic limit \cite{Barmettler_2009,Barmettler_2010}. The inability of a finite system to relax has two origins. Note that a many-body state can have a small order parameter if it is a linear combination of symmetry-related Fock states like $\ket{1010...}$ and $\ket{0101...}$. We have verified that the projection of $ e^{-i\hat{H}\tau}\ket{1010...}$ onto $\ket{0101...}$ is negligible during the time evolution for considered system sizes. Nevertheless, the maximal value of the projection increases with $L$. Furthermore, a larger dimension of the Hilbert space provides a larger number of symmetry-related Fock states. Although a finite system cannot completely get rid of its charge density wave order, it is expected to follow the exponentially-slow decay for longer times $\tau_\text{rev}$, the more atoms it has.

On the other hand, a many-body state can have a small order parameter if it is a linear-combination of Fock states with a small order parameter. These are typically related to a high probability of finding neighbouring HCBs, which in turn leads to a large expectation value of interaction energy. The considered systems of interacting HCBs in the absence of the laser pulse are closed and support the conservation of energy. Therefore, $\Psi\left(\tau\right)$ is expected to be composed of Fock states with expectation values of energies close to zero. This explains why the order parameter, which is not able to vanish in the unperturbed finite system, rapidly decreases to zero after an excitation by a laser pulse, i.e., after an absorption of a large amount of energy. A somewhat similar revival of nearest-neighbour correlations has been observed in Ref.~\cite{Mishra_2019}, in which the finite system dynamics has been governed by the XY Heisenberg model with a square-wave magnetic field. In this study, $\Psi\left(0\right)$ corresponds to a low-temperature state of the initial Hamiltonian.

Note that even for very short laser pulses with $A\gtrsim 1$ and $\sigma<1$, the order parameter is significantly decreased during the pulse duration (Fig.~\ref{fig1} and Fig.~\ref{fig2}). It seems that the charge density wave or N{\'e}el order is not completely lost when $\omega=V$ (Fig.~\ref{fig1}). A partial translation of HCBs or a partial reversal of spins, which is characterized by $\bar{\eta}<0$, is supported for long times. We elaborate on this in the next section. We have also investigated the time evolution of the order parameter in the case of periodic driving, and we have obtained a suppression that is reproduced for pulses with $\sigma\gtrsim 1$. Besides the suppression, even the overall time evolution of the order parameter under the periodic driving is qualitative similar as for a pulse with $\sigma=5$. It closely resembles $\eta\left(\tau\right)$ of the unperturbed system below the quantum phase transition with $V<2$ \cite{Barmettler_2009,Barmettler_2010}, i.e., the electric field effectively suppresses the nearest-neighbour interactions.

On the other hand, for very short laser pulses with $A<1$ and $\sigma< 1$, the order parameter is reduced to $\bar{\eta}> 0$ during the pulse duration (Fig.~\ref{fig3}(a)). The following time evolution mimics the unperturbed one with non-vannishing oscillations around a time-independent average $\bar{\eta}> 0$, which decreases with an increasing pulse duration. The presented results signal a possibility of a dynamic control of the order parameter (the staggered magnetization in the XXZ Heisenberg chain). Similar possibility has been previously considered in a different system, i.e., the periodically driven Fermi-Hubbard model on the Bethe lattice in the limit of infinite coordination number \cite{Mendoza_2017}. We adopt one of the protocols proposed in Ref.~\cite{Mendoza_2017}, and adapt it to a different setup, see Fig.~\ref{fig3}(b). We first excite the system with a laser pulse with $A=0.5$, $\omega=V$ and $\tau_0 = 6$. Subsequently, we excite the system with another laser pulse with $A=2.4$, $\omega\gg V$ and $\tau_0=27$, which slows down the dynamics of HCBs. Consequently, a plateau appears in the time evolution of the order parameter, whereby its value and size are controlled by the duration of the first and the second pulse, respectively. Note that the use of the two pulses instead of a periodic electric field, as shown in Ref.~\cite{Mendoza_2017}, may simplify the introduction of this protocol to experiments with ultra-cold atoms.

\section{Results: Non-thermal long-lived states}

We now focus on the most unexpected result shown first in Fig.~\ref{fig1}(a), i.e., the order parameter after a resonant-like pumping with $A\gtrsim 1$, $\omega=V$ and $\sigma<1$ appears to be non-thermal. It should be emphasized that the protocols previously introduced in similar systems did not create non-thermal states with negative order parameters (see for example Ref.~\cite{Collura_2020,Balzer_2015,Mendoza_2017,Barmettler_2009,Cramer_2008}). In our protocol the initial charge density wave order is partially kept, although with HCBs translated by one lattice site. We confirm that $\eta\left(\tau\right)<0$ is maintained for long times, longer than typical times $\frac{\hbar}{V}$ and $\frac{\hbar}{t}$ given by model parameters (see Fig.~\ref{fig4}(a) for an exemplary time evolution with $\tau\leq 50$ after a pumping with $\sigma=0.5$). The gradual progression of the order parameter towards zero up to oscillations is quantitatively described as
\begin{equation}
\label{eq2}
\eta\left(\tau\right)= - B e^{-\tau/\tau_r}.
\end{equation}
The parameter $B$ is a monotonically decreasing function of duration, and it approaches zero when $\sigma\rightarrow 1.0$ (see the inset of Fig.~\ref{fig4}(a)). The relaxation time $\tau_r$ is a concave function with a maximum near $\sigma\approx 0.4$ (or $0.5$ depending on a system size). Furthermore, $\tau_r > 20.0$ for all $\sigma< 1$. We have established that the non-thermal states are stable against minor changes in the interaction strength $V$ (Fig.~\ref{fig4}(c)), and amplitude $A$ (Fig.~\ref{fig4}(d)). As a result, we propose the existence of long-lived non-thermal states in a finite one-dimensional system of HCBs generated by laser pulses with amplitudes $A\gtrsim 1$, a resonant-like frequency $\omega=V\gtrsim 7$, and durations $\sigma< 1$. In the appendix A, we explain the assumption apparent in Eq.~\ref{eq2} that $\lim_{\tau\rightarrow\infty}\eta\left(\tau\right)=0$ in finite systems.

We have also examined long-lived non-thermal states in systems with different numbers of lattice sites $L\in\left\{14,16,...,26\right\}$ (Fig.~\ref{fig4}(b)). We have found that the parameter $B$ weakly depends on $L$ (not shown). On the other hand, the relaxation time increases with $L$ (the inset of Fig.~\ref{fig4}(b)). We are currently not able to provide a function $\tau_r\left(L\right)$, which would determine the behaviour of the relaxation time in the thermodynamic limit. Note that already for $L=20$, the relaxation time exceeds the maximal time considered in numerical calculations and, so, the accuracy of the fit $\eta\left(\tau\right)=-Be^{-\tau/\tau_r}$ decreases with $L\geq 20$. It should also be highlighted that $\tau_r$ of a steady-like state, and its potential saturation or divergence in the thermodynamic limit, seems to depend on the time of excitation with the laser pulse (see Fig.~\ref{fig4}(b) for the black curve corresponding to $L=24$ and $\tau_0=6$, as well as the rainbow curves corresponding to $L\in\left\{14,16,...,26\right\}$ and $\tau_0=1$).

In view of all findings, we speculate that the long-lived non-thermal states exist in the thermodynamic limit. Whether the charge density wave order decays to zero or not for $\tau\rightarrow\infty$ and $L \rightarrow\infty$ remains, however, an open question. If it does not decay, the long-lived non-thermal states have to be inextricably related to the itegrability of the considered model, so that the inclusion of next-nearest-neighbour terms in the Hamiltonian from Eq.~\ref{eq1} should prevent their formation. The inclusion of these terms results in a pronounced decrease of the charge density wave order in the investigated finite systems, as explained in the next section.

Let us point out that, despite the coincidence of names, the observed long-lived non-thermal states cannot be identified with the widely-studied prethermalized states \cite{Moeckel_2008}. The prethermalized states and the related prethermalized plateaus of observables are characterized by finite lifetimes even in large systems. They are established in nearly but not exactly integrable models, and can be explained with the existence of approximate constants of motion \cite{Kollar_2011}.
\begin{figure}
\centering
\includegraphics[width=0.8\columnwidth]{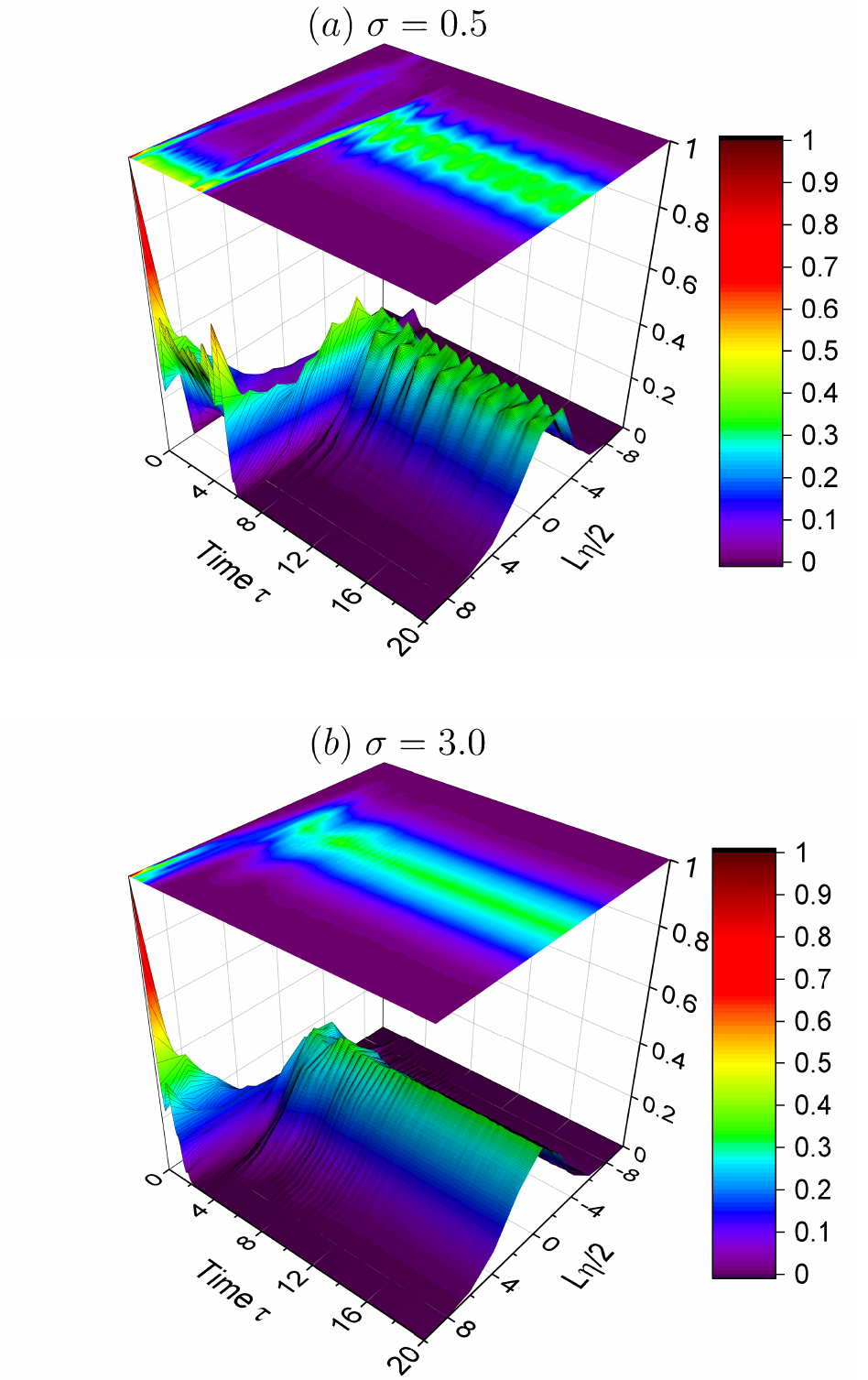}
\caption{The probability distribution of the staggered magnetization  $P\left(m\right)$ as a function of time $\tau\in[0,...,20]$ for a system with $L=20$ and (a) a short pulse with $\sigma=0.5$ and (b) a long pulse with $\sigma=3.0$. The amplitude is $A=2.5$. The initial probability distribution is peaked at $m=L/2$, while the long-time probability distribution is a Gaussian with a mean staggered magnetization equal zero for (b), or shifted towards negative values for (a).}
\label{fig5}
\end{figure}
\begin{figure}
\centering
\includegraphics[width=0.8\columnwidth]{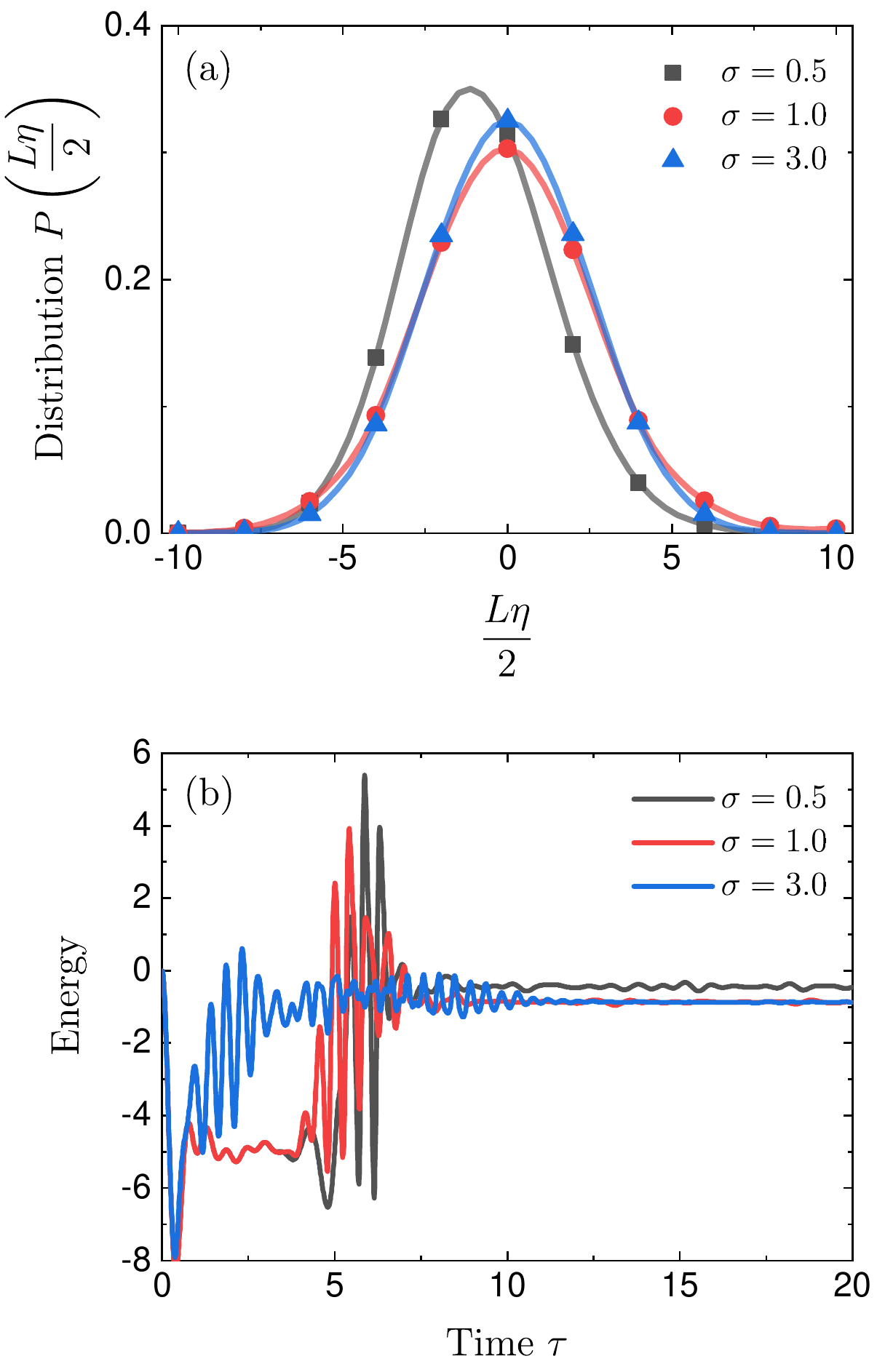}
\caption{(a) The probability distribution of the staggered magnetization  $P\left(m\right)$ averaged over times $\tau\in[10,...,20]$. Note that for a perfect charge density wave state, i.e., $\Psi\left(0\right)$, the probability distribution is given by $P\left(m\right)=\delta\left(\frac{L}{2}-m\right)$. Points are numerical data, while lines are fitted Gaussian functions. The probability distribution is apparently different for non-thermal long-time states, e.g., it lacks the symmetry in $m$. (b) The time evolution of a kinetic energy of HCBs. Note that the kinetic energy of non-thermal long-time states saturates closer to zero. A system with $L=20$ is considered.}
\label{fig6}
\end{figure}

Before moving to the next section, let us characterize the observed long-lived non-thermal states in more details. Following Ref.~\cite{Collura_2020}, we introduce the probability distribution of the staggered magnetization $m=L\eta/2$,
\begin{equation}
P\left(m\right)= \bra{\Psi\left(\tau\right)} \delta\left(\sum\nolimits_{j=1}^{L} \left(-1\right)^j c_j^\dagger c_j - m\right) \ket{\Psi\left(\tau\right)}
\end{equation}
which satisfies $\sum_{m} P\left(m\right)=1$ and is defined in points $m = -L/2+2n$ for integer $n\in[0,...,L/2]$. As presented in Fig.~\ref{fig5}, the probability distribution for longer pulses with $\sigma\gtrsim 1$ is approximately Gaussian at all times $\tau>\tau_0$. Therefore, the time evolution of $P\left(m\right)$ is qualitatively similar to the one in the unperturbed system below the phase transition (see the appendix~C and Ref.~\cite{Collura_2020} for results in a finite system and the thermodynamic limit, respectively). The probability distribution in the unperturbed system above the phase transition has a large peak near $m=\frac{L}{2}$ and exhibits a pronounced oscillatory behaviour or even-odd structure in $m$ for short times. It remains strongly non-Gaussian even for longer times \cite{Collura_2020}. Moreover, we observe that for longer pulses with $\sigma\gtrsim 1$ the mean staggered magnetization initially oscillates between negative and positive values, but quickly becomes equal to zero. As a result, the long-time states support the spin-rotational symmetry of the XXZ Heisenberg chain. Although the probability distribution for shorter pulses with $\sigma<1$ is also approximately Gaussian at all times $\tau>\tau_0$, the mean staggered magnetization is shifted towards negative values. The difference between thermal and non-thermal long-time states is more apparent in Fig.~\ref{fig6}(a), where $P\left(m\right)$ averaged over times $\tau\in\left[10,...,20\right]$ are displayed. The difference is also revealed in the kinetic energy of HCBs, which saturates closer to zero when the system is excited with a shorter pulse with $\sigma<1$ (Fig.~\ref{fig6}(b)). In the appendix B, we show the quasi-momentum distribution which further confirms the existence of residual correlations between HCBs for $\sigma<1$.

\section{Results: Integrability breaking}
\begin{figure}
\centering
\includegraphics[width=0.8\columnwidth]{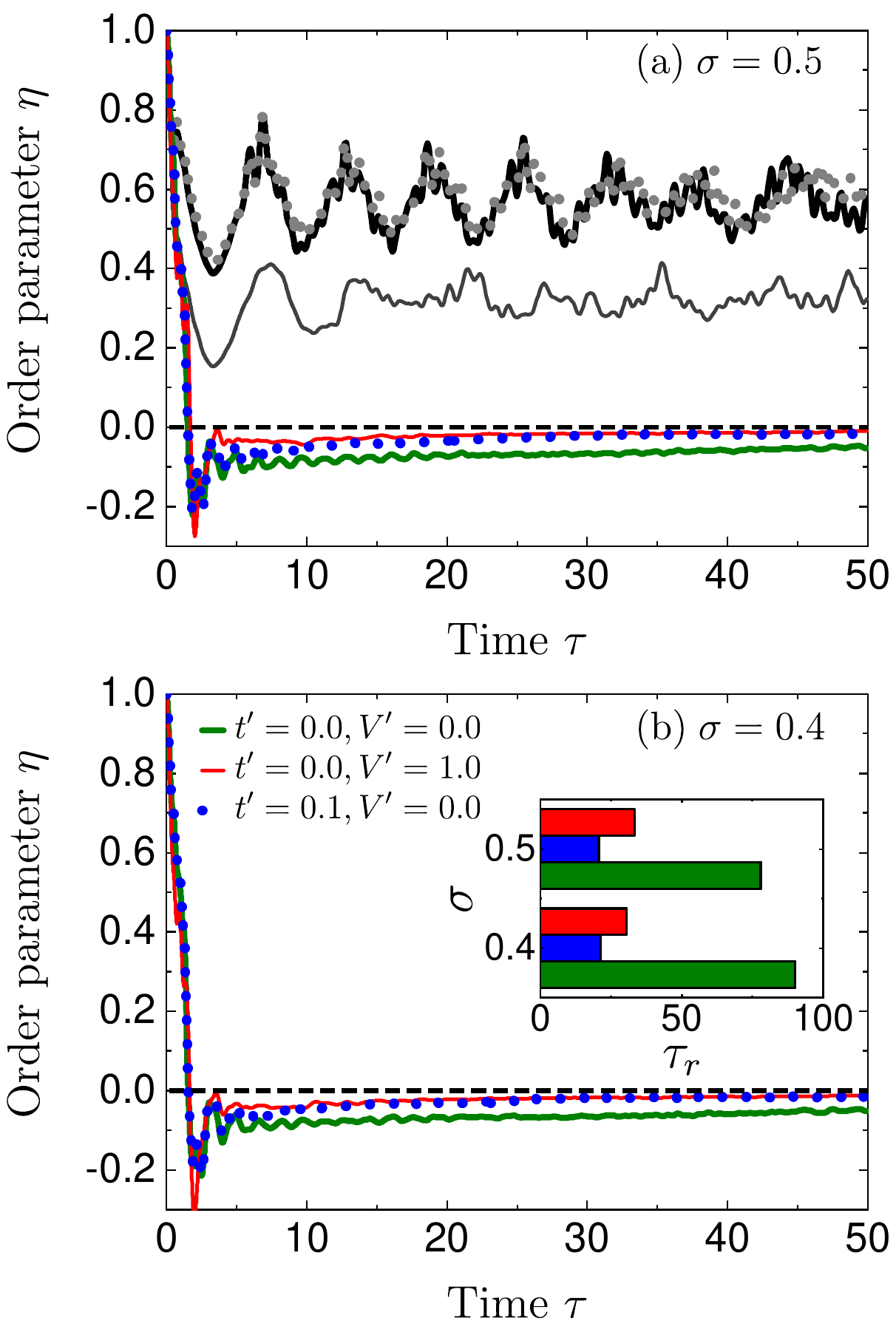}
\caption{The comparison of long-lived non-thermal states in a finite integrable system with $V^{'}=t^{'}=0.0$ (green thick lines), and after the integrability breaking with next-nearest-neighbour interactions $V^{'}=1.0$ (red thin lines) as well as a next-nearest-neighbour hopping $t^{'}=0.1$ (blue dotted lines). Laser pulses with amplitude $A=2.5$, frequency $\omega=V$, and durations (a) $\sigma=0.5$ and (b) $\sigma=0.4$ are studied. We chose to excite the system at $\tau_0=1$. The black thick, grey thin and light grey dotted curves correspond to $\eta\left(\tau\right)$ without pumping for $V^{'}=t^{'}=0.0$, $V^{'}\neq 0$ and $t^{'}\neq 0$, respectively. It is apparent that the relaxation of order parameter towards zero is accelerated when the next-nearest-neighbour terms are included in the Hamiltonian from Eq.~\ref{eq1}. The latter conclusion is supported by the inset, which presents relaxation times $\tau_r$ for all considered scenarios. The colour of the bar in the inset corresponds to the colour of the curve in the plot. The legend presented in (a) is valid for (b) as well.}
\label{fig7}
\end{figure}
Integrable models are characterized by an extensive number of local conserved quantities, and have different properties than quantum-chaotic models \cite{Grabowski_1995,Mierzejewski_2015,Ilievski_2016}. Their relaxation dynamics is not governed by the Eigenstate Thermalization Hypothesis and, therefore, their behaviour is unique when taken far from equilibrium. For example, the infinite-time averages of expectation values can disagree with predictions of the grand-canonical ensemble. Instead, they are described by the Generalized Gibbs Ensemble, in which not only average energy and particle number but also other constants of motion are fixed \cite{Cassidy_2011,Ilievski_2015,Essler_2016,Vidmar_2016}. The latter accounts for the fact that the time evolution of an arbitrary state in an integrable system is performed in a restricted Hilbert space.

Non-thermal states evidenced in the paper are long lived, and their relaxation times increase with the number of lattice sites. We were not able to rule out the possibility that the charge density wave order remains non-zero in the long-time and thermodynamic limits. If this is the case, the observation of non-thermal states is possible due to the integrability of XXZ Heisenberg chain. Simultaneously, the observed relaxation times should significantly diminish when the system is taken away from its integrability point. We consider two such scenarios. In the first one we introduce next-nearest-neighbour interactions $V^{'}\sum_{j=1}^{L} n_j n_{j+2}$, while in the second one we enable a next-nearest-neighbour hopping $t^{'} \sum_{j=1}^L \left(e^{i2A\left(\tau\right)} c_j^\dagger c_{j+2} + \text{H.c.} \right)$. The time evolution of the order parameter after the inclusion of integrability-breaking terms in the Hamiltonian from Eq.~\ref{eq1} is presented in Fig.~\ref{fig5}. In both scenarios, the charge density wave order is substantially reduced. We find exceptionally surprising that the relaxation time is significantly reduced in the case when $t^{'}=0.1$ (see the green and blue bars in the inset of Fig.~\ref{fig5}(b)), even though the time evolution of the order parameter in the system without pumping remains almost the same as for $t^{'}=0.0$ (see the black thick and light grey dotted curves in Fig  \ref{fig5}(a)).

\section{Concluding remarks}

In this paper we have studied a unitary time evolution of the state with alternately filled and unfilled sites in a finite one-dimensional system of interacting HCBs under the influence of a spatially-homogenous and time-dependent vector potential that mimics a laser pulse. We have restricted our investigations to strongly-correlated systems, i.e., the gapped phase with $V\gg 2$. We have demonstrated that it is possible to dynamically control the order parameter when the excitation by a laser pulse with amplitude $A<1$ and frequency $\omega=V$ is followed by the excitation by a laser pulse with amplitude $A=2.4$ and frequency $\omega\gg V$, which slows down the dynamics of HCBs. On the contrary, the order parameter is significantly diminished during the pulse duration when $A\gtrsim 1$. We have also observed an interesting relation between the absorbed energy and the moment of excitation $\tau_0$. Since the considered system is not in an energy eigenstate at the beginning of the time evolution, it evolves in time before experiencing a laser pulse, and its response depends on $\tau_0$. This observation can be important for pump-probe-like
experiments.

Moreover, we have established a protocol in which the system becomes trapped in a long-lived non-thermal state characterized by an order parameter with a reversed sign. In the spin picture, this is consistent with a partially preserved N{\'e}el order with spin orientation reversed. In the language of HCBs, the charge density wave is partially preserved but translated by one latice site. Furthermore, the corresponding relaxation time is large in comparison to typical times given by model parameters, e.g., $\hbar/t$ and $\hbar/V$, and it even grows with the number of lattice sites. Our results suggest that the long-lived non-thermal states exist in the thermodynamic limit. However, they are inconclusive whether the charge density wave order decays to zero or not when $\tau\rightarrow\infty$ and $L\rightarrow\infty$. A protocol in which a finite one-dimensional system of interacting HCBs remains for a long time after a perturbation is gone in a non-thermal state with partially preserved nearest neighbour correlations of an initial state has not been proposed before. Nevertheless, it is somewhat related to the protocol in which a finite one-dimensional system of interacting HCBs with filled all sites in the centre and empty all sites near the edges experiences a sudden quench of a confining potential and undergoes a sudden expansion along a chain. In the latter scenario, long-lived quasi-condensed states have been observed \cite{Vidmar_2015_hcb}.

Since the existence of non-thermal states in the long-time and thermodynamic limits is expected only in integrable models (e.g., our model before and after it is subjected to a laser pulse with a Gaussian profile), we have repeated the time evolution with integrability breaking terms included in the Hamiltonian (i.e., next-nearest-neighbour hopping $t^{'}$ and interactions $V^{'}$). We have observed a much faster decrease of the magnitude of the order parameter. This is particularly unexpected in the scenario with a non-zero hopping between next-nearest sites $t^{'}=0.1$, since the time evolution of the order parameter without pumping is almost identical as $\eta\left(\tau\right)$ for $t^{'}=0$.

\section*{ACKNOWLEDGMENTS}
We thank L. Vidmar for comments and insightful questions about our results. This work was supported by the Slovenian Research Agency (ARRS), Research Core Fundings Grant No. P1-0044. J.~B. acknowledges the support from the Centre for
Integrated Nanotechnologies, a U.S. Department of Energy, Office of Basic Energy Sciences user facility.


\bibliographystyle{biblev1}
\bibliography{references}

\raggedbottom
\section{Appendix A}
\begin{figure}[H]
\centering
\includegraphics[width=0.8\columnwidth]{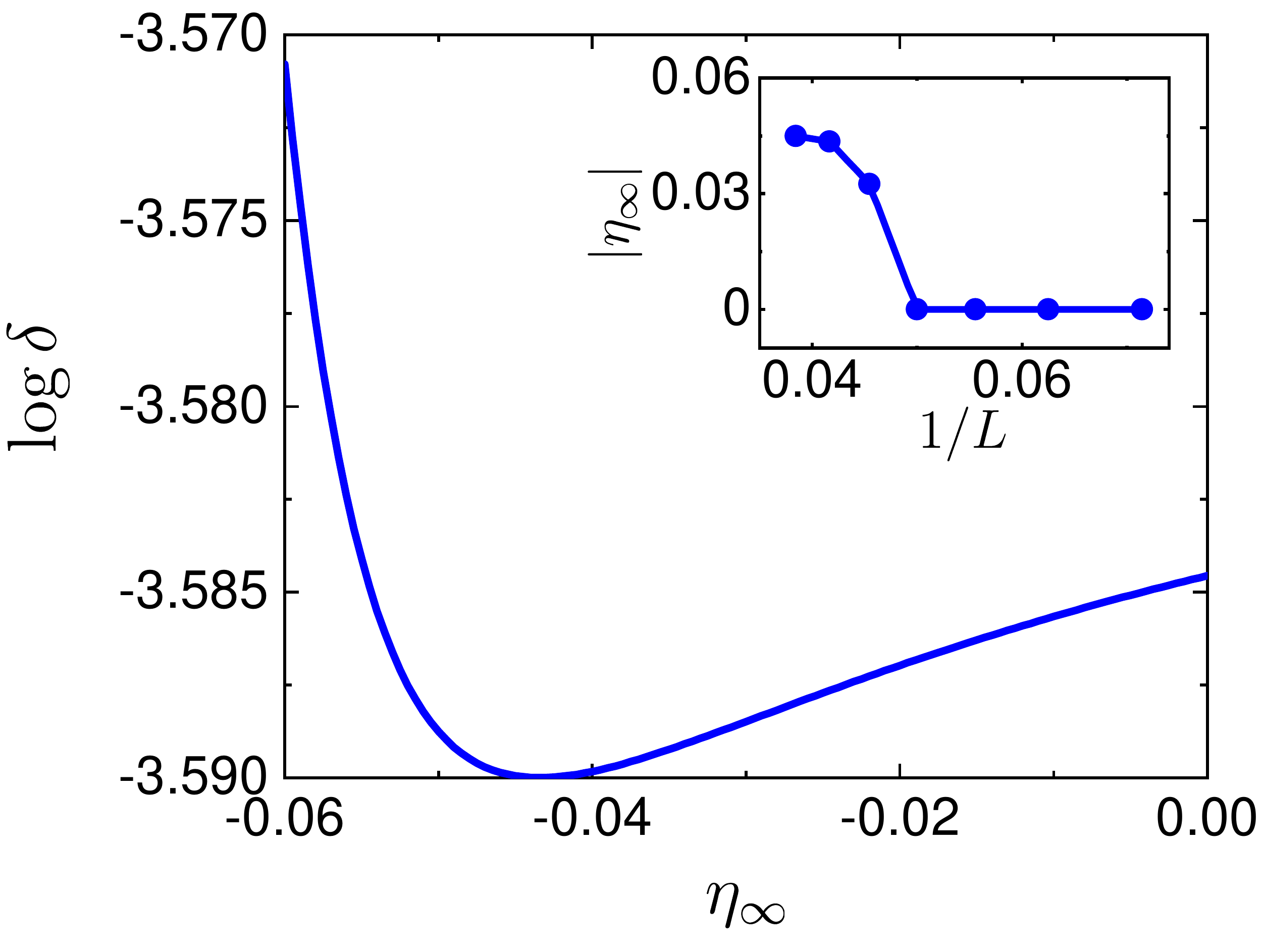}
\caption{The quality of the fit $\log\delta$ versus the infinite-time value of order parameter $\eta_\infty$ (see the main text for explanations). The considered system of HCBs has $L=24$ sites and it is perturbed by a laser pulse with amplitude $A=2.5$, frequency $\omega=V$, and duration $\sigma=0.5$. The inset shows how $|\eta_\infty|$ corresponding the minimal value of $\log\delta$ increases with $L$.}
\label{figA1}
\end{figure}
In the main text we have assumed that the infinite-time average of the order parameter, $\eta_\infty$, is zero for long-lived non-thermal states which are realized in finite systems. The most likely value of $\eta_\infty$ can be determined from the quality of the fit $f\left(\tau\right)=- B e^{-\tau/\tau_r}+\eta_\infty$ with two free parameters $\left\{A,\tau_r\right\}$ and one fixed parameter $\eta_\infty$. We have defined the accuracy of this fit as $\delta^2=\sum_{j=1}^{N} \left(f\left(\tau_j\right)-\eta\left(\tau_j\right)\right)^2/N$. Although the most likely value of $\eta_\infty$ is negligible for small systems, it becomes negative for $L> 20$ (the inset of Fig.~\ref{figA1}). Nevertheless, the differences in $\log\delta$ in the interval $\eta_\infty\in\left[-0.6,0.0\right]$ are very small (Fig.~\ref{figA1}) and, so, the problem is somewhat ill-defined and the assumption that $\eta_\infty$ is non-zero seems insufficiently justified. Therefore, we have adopted the worst case scenario in which $\eta_\infty=0$. A non-zero infinite-time average of the order parameter in the thermodynamic limit can still result from a diverging relaxation time $\tau_r\rightarrow\infty$ and not from $\eta_\infty<0$.

\section{Appendix B}
\begin{figure}[th!]
\centering
\includegraphics[width=0.8\columnwidth]{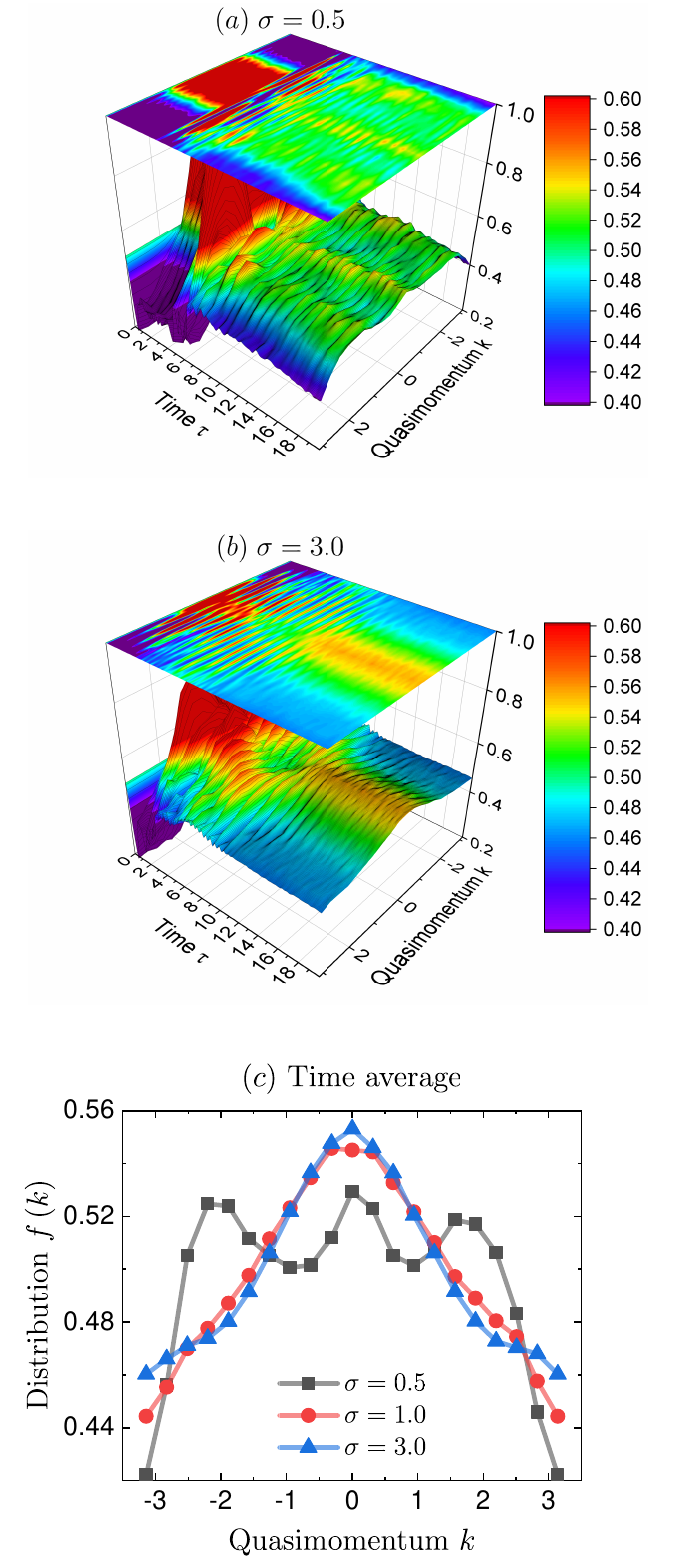}
\caption{The quasi-momentum distribution $f\left(k\right)$ as a function of time $\tau\in[0,...,20]$ for a system with $L=20$ and (a) a short laser pulse with $\sigma=0.5$ and (b) a long laser pulse with $\sigma=3.0$. Plots obtained after averaging over times $\tau\in[10,...,20]$ are presented in (c). See a single-peak structure for long pulses and a three-peak structure for short pulses. The latter is qualitatively similar to the qusi-momentum distribution of a ground state of HCBs placed in a superlattice potential $V_{ext}=V_0\sum_{j=1}^{L} \cos\left(\pi j\right) c_j^\dagger c_j$. The amplitude is $A=2.5$.}
\label{figA2}
\end{figure}

In Fig.~\ref{figA2}, we demonstrate the quasi-momentum distribution normalized to the number of particles,
\begin{equation}
f\left(k\right) = \bra{\Psi\left(\tau\right)}\frac{1}{L} \sum\nolimits_{m\neq n} e^{-ik\left(m-n\right)}c_m^\dagger c_n \ket{\Psi\left(\tau\right)}
\end{equation}
with $k\in[-\pi,\pi)$. The initially flat $f\left(k\right)$ develops a single maximum in $k=0$, which is divided by a laser pulse into two maxima placed symmetrically around $k=0$. For long times $\tau>10$, for which the influence of an electric field is already minimal, the quasi-momentum distribution stabilizes. Although slight oscillations are still visible for $\tau\approx 20$. For long pulses with $\sigma\gtrsim 1$, the long-time distribution has a single peak in $k=0$. For short pulses with $\sigma>1$, the long-time distribution has a peak in $k=0$ but also retains two maxima near $\pm\pi/2$. It should be noted that a qualitatively similar three-peak structure is accomplished for a ground state of HCBs placed in a superlattice potential $V_{ext}=V_0\sum_{j=1}^{L} \cos\left(\pi j\right) c_j^\dagger c_j$ \cite{Rigol_2007}. This can be associated with a residual charge density wave order.

\section{Appendix C}
\begin{figure}[H]
\centering
\includegraphics[width=0.8\columnwidth]{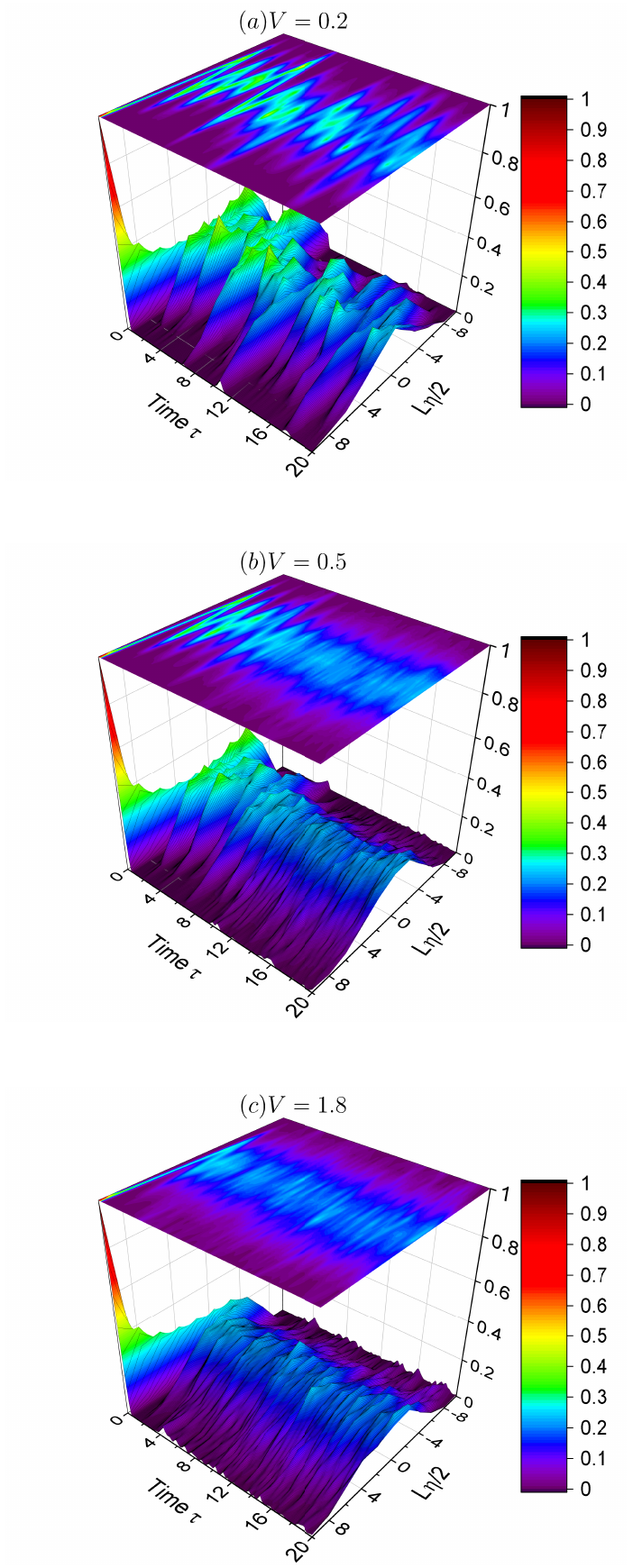}
\caption{The probability distribution of the staggered magnetization $P\left(m\right)$ as a function of time $\tau\in[0,...,20]$. The systems of interacting HCBs with $L=20$, which is not perturbed by any laser pulse, below the phase transition with (a) $V=0.1$, (b) $V=0.5$ and (c) $V=1.8$ is considered. Note that the initial probability distribution is given by $P\left(m\right) = \delta\left(\frac{L}{2}-m\right)$.}
\label{figA3}
\end{figure}

The probability distribution of the staggered magnetization $m=L\eta/2$,
\begin{equation}
P\left(m\right)= \bra{\Psi\left(\tau\right)} \delta\left(\sum\nolimits_{j=1}^{L} \left(-1\right)^j c_j^\dagger c_j - m\right) \ket{\Psi\left(\tau\right)}
\end{equation}
satisfies $\sum_{m} P\left(m\right)=1$ and is defined in points $m = -L/2+2n$ for integer $n\in[0,...,L/2]$. In Fig.~\ref{figA3}, we present the time evolution of $P\left(m\right)$ in the system of interacting HCBs below the phase transition, which is not perturbed by a laser pulse. Initially the probability distribution is non-zero only in one point, but after a short time $P\left(m\right)$ spreads in $m$ and becomes satisfactorily described by a Gaussian function with an interaction-dependent standard deviation.


\end{document}